\documentclass[%
 reprint,
 amsmath,amssymb,
 aps,
pra,
]{revtex4-2}

\usepackage{graphicx}% Include figure files
\usepackage{dcolumn}% Align table columns on decimal point
\usepackage{bm}% bold math
\usepackage{hyperref}% add hypertext capabilities
\begin{document}

\preprint{APS/123-QED}

\title{Laser propagation in Rindler accelerated reference 
	frame based on matrix optics
}% Force line breaks with \\

\author{Weifeng Ding}
 %\altaffiliation[Also at ]{Zhejiang Province Key Laboratory of Quantum Technology and Device, Department of Physics, Zhejiang University, Hangzhou 310027, China.}%Lines break automatically or can be forced with \\
\author{Zhaoying Wang}%
 \email{zhaoyingwang@zju.edu.cn}
\affiliation{%
 Zhejiang Province Key Laboratory of Quantum Technology and Device, Department of Physics, Zhejiang University, Hangzhou 310027, China
}%


%\affiliation{


\date{\today}% It is always \today, today,

\begin{abstract}
The Rindler space-time describing a series of accelerating observers is Ricci flat, but it still has novel optical effects. In the case of WKB approximation, we derive the light paths in the Rindler frame based on the covariant wave equation and geodesic equations. Then, we use ABCD matrix optics method to explore the propagation characteristics of Rindler frame, thus link three different optical transformation scenes (geometry, gravity and vacuum refractive index) together. Moreover, the propagation characteristics of hollow beam in Rindler space-time are described analytically. In the longitudinal direction, we demonstrate the shift and stretch effects of the dark spot of a beam, while the transverse spot size is proved to be convergence in the accelerated system, and the wavefront curvature can tend a constant twice the acceleration at the far field. Those characteristics are quite different from the ones in the flat space-time. Based on these calculations, we simply demonstrate the position uncertain relationship between the transverse beam size and the momentum, which surprisingly coincides with the derivation of quantization. We hope that we can provide one simple method to analyze the beam propagation in the accelerated frame.
%\begin{description}
%\end{description}
\end{abstract}

\maketitle

%\tableofcontents

\section{Introduction}

In recent years, the research devoted to applying general relativity (GR) concept to optics has become increasingly rich, from the intuitive curved surface \cite{RN1,RN2,RN3,RN4} to the abstract curved space-time \cite{RN5,RN6,RN7}. Compared with the optical theoretical research \cite{RN3,RN7,RN8}  and analog experiments \cite{RN6,RN9,RN10} of Schwarzschild black holes, it is meaningful to study a more general and simple space-time background: the accelerated reference frame, research  based on which for optics appears to be rare until now.

For flat, gravitation-free space-time, we can describe it in terms of Minkowski space-time, its line element is satisfied as 
$d{s^2} =  - {c^2}d{T^2} + d{X^2} + d{Y^2} + d{Z^2}$, where $T,{\text{ }}X,{\text{ }}Y,{\text{ }}Z$ are the space-time coordinates, and $c$ is the speed of light. According to Einstein, the curvature of space-time leads to the bending of light, which was first observed by Eddington in 1919. According to Einstein's equivalence principle (EEP) \cite{RN11}, we can locally regard an accelerated reference frame as an inverted gravitational field ($\kappa=-a$). There are a lot of wonderful research due to the bending of light in gravitational fields such as gravitational lensing \cite{RN7,RN12,RN13}, Redshift effect \cite{RN14,RN15}, black hole shadow \cite{RN16} and so on. Correspondingly, more attention should be paid to the propagation characteristics of light in the accelerated reference frame. W. Rindler studied a coordinate transformation between an accelerated reference frame and a flat Minkowski space-time \cite{RN17}, where the inherent acceleration is constant. When the accelerated reference frame is only along the z axis, the coordinate transformation can be given by $T\left( {t,z} \right) = \left( {{{{e^{az}}} \mathord{\left/
			{\vphantom {{{e^{az}}} a}} \right.
			\kern-\nulldelimiterspace} a}} \right)\sinh at$, $Z\left( {t,z} \right) = \left( {{{{e^{az}}} \mathord{\left/
			{\vphantom {{{e^{az}}} a}} \right.
			\kern-\nulldelimiterspace} a}} \right)\cosh at$. This is known as the Rindler transform, where the capital $T$ and $Z$ represent the Minkowski coordinates in the inertial frame, $t$ and $z$ represent the coordinates in accelerated reference frame. Here, the speed of light is set as $c=1$, so the acceleration $a$ in the formula is the normalized acceleration $a/c^{2}$.\\
\indent By substituting the above coordinate transformation into a four-dimensional Minkowski space-time, the conformal form of line element in the Rindler accelerated reference frame is given as $d{s^2} = {e^{2az}}\left( { - d{t^2} + d{z^2}} \right) + d{y^2} + d{x^2}$. It is known that the Rindler coordinates describe the comoving frame of the accelerating observer, which locally equivalent to Minkowski space-time and therefore has no Ricci curvature. However, the accelerated reference frame brings about some intriguing properties in electromagnetism field such as Unruh effect \cite{RN18}, generalized uncertainty principle \cite{RN19} (GUP) and non-quantum thermal fluctuations \cite{RN20}. So, the study of optical transmission characteristics in Rindler space-time is novel and worth digging.\\
\indent In this paper, we deconstruct the optical transmission characteristics of Rindler reference frame from the perspective of pure optics and regard it as a special optical transmission system. We introduce the processing method of matrix optics into the space-time discussion for the first time, attempt to simplify the problem of light transmission under the metric of Rindler reference frame and construct the relationship between the metric and ABCD matrix. Through our calculation, we find that the accelerated reference frame can also have its corresponding vacuum refractive index, which expands the scope of application of vacuum refractive index theory \cite{RN21,RN22,RN23}, and it implies that the geometric description and space-time background description of optical propagation are closely related to the vacuum refractive index. On this basis, we analyze and study the propagation characteristics of Gaussian hollow beam (HGB) in the Rindler accelerated reference frame. It is clarified that the spatiotemporal background influences the intrinsic properties of the beam, which is different from the geometric bending \cite{RN12,RN24}. This research tells that even in the locally flat Rindler reference frame, the propagation properties of the beam will be changed to some extent.

\section{Matrix optics of accelerated frame}

The ABCD matrix method is powerful to deal with the axisymmetric optical system because of its simplicity and cleanness. According to the hypothesis of GR, that light travels on null geodesics in space-time which means $d{s^2} = 0$.Using the metric, we can simply obtain the following expression \cite{RN25}:
\begin{equation}
	\label{metri}
	\left.
	d{s^2} = {g_{\mu \nu }}d{\kappa ^\mu }d{\kappa ^\nu } = 0,
	\right.
\end{equation}
where the indices $\mu ,{\text{ }}\nu$ run over $0,1,2$ and satisfy Einstein's summation rule. If we consider light traveling along the z-axis, in the same direction as the acceleration of Rindler observer, ${\kappa ^0},{\text{ }}{\kappa ^1},{\text{ }}{\kappa ^2}$ in Eq. \eqref{metri} represent $t,{\text{ }}z,{\text{ }}r{\text{ }}$, respectively.
\begin{equation}
	\label{guv}
	\left.
	{g_{\mu \nu }} = diag\left( { - {e^{2az}},{e^{2az}},1} \right),
	\right.
\end{equation}
stands for the Riemann metric.

The null geodesic condition is not sufficient to  obtain the path of the light, and therefore we need to consider the wave function of the light in the four-dimensional space-time background. In this case, we use massless scalar wave equation to consider optical propagation \cite{RN14}.
\begin{equation}
	\label{wave1}
	\left.
	\square \phi  + R\xi \phi  = 0,
		\right.
\end{equation}
Where $R$ stands for the Ricci scalar curvature, which is 0 here. $\square$ is the covariant d 'Alembert operator $\square=\frac{1}{{\sqrt g }}{\partial _\mu }\sqrt g {g^{\mu \nu }}{\partial _\nu }$, where $g$ is the determinant of the metric. $\phi$ can be separated variables by $\phi  = {\phi _s}\left( {r,z} \right){\phi _t}\left( t \right)$. For our diagonal metric, the wave equation of spatial part has the following simple form.
\begin{equation}
	\label{4}
	\left.
{e^{ - 2az}}{\partial _r}^2{\phi _s} + {\partial _z}^2{\phi _s} + 4a{\partial _z}{\phi _s} + {\zeta ^2}{\phi _s} = 0,
	\right.
\end{equation}
here $\zeta$ originates from the separation of variables, which acts as a propagation constant. 
We choose the wave function of the form ${\phi _s} = U\exp (ikL)$, under the WKB approximation, dropping the second derivative term of $U$, and the real part and the imaginary part are equal to zero,  so we have the following two equations.
\begin{equation}
	\label{5}
	\left.
{\text{Re:}}{\left( {{\partial _z}L} \right)^2} + {e^{ - 2az}}{\left( {{\partial _r}L} \right)^2} - 1 = 0,
	\right.
\end{equation}
\begin{equation}
	\label{6}
	\left.
\operatorname{Im} :2{\partial _r}U{\partial _r}L + U\left[ {{\partial _r}^2L + {e^{2az}}\left( {4a{\partial _z}L + {\partial _z}^2L} \right)} \right] = 0.
\right.
\end{equation}
In flat space, the physical meaning of $L$ is eikonal \cite{RN3,RN26}, and we're dealing with a vacuum background, so it's essentially the length of the light that travels, and if we define the input and output field, using Green's function we have:
\begin{equation}
	\label{7}
	\left.
{\phi _{out}} = {\phi _{in}} \otimes {e^{ikL}}.
\right.
\end{equation}
For the paraxial case under weak acceleration, according to the knowledge of differential geometry, a geodesic is a line with zero geodesic curvature, i.e.:
\begin{equation}
	\label{8}
	\left.
\frac{{{d^2}{\kappa ^\mu }}}{{d{\tau ^2}}} + \Gamma _{\nu \sigma }^\mu \frac{{d{\kappa ^\nu }}}{{d\tau }}\frac{{d{\kappa ^\sigma }}}{{d\tau }} = 0.
\right.
\end{equation}
Where $\Gamma _{\nu \sigma }^\mu \!=\!\frac{1}{2}{g^{k\mu }}\left( {\frac{{\partial {g_{\nu k}}}}{{\partial {\kappa ^\sigma }}} + \frac{{\partial {g_{\sigma k}}}}{{\partial {\kappa ^\nu }}} - \frac{{\partial {g_{\nu \sigma }}}}{{\partial {\kappa ^k}}}} \right)$ called the Christoffel symbol which describes the affine contact of coordinate. $\tau$ can be any affine parameter, plug in the Rindler metric, and consider the null geodesic condition, then we have these equations:
\begin{equation}
	\label{9}
	\left.
\left\{ {\begin{array}{*{20}{ll}}
		{\frac{{{d^2}t}}{{d{\tau ^2}}} + 2a\frac{{dt}}{{d\tau }}\frac{{dz}}{{d\tau }} = 0}\vspace{6pt} \\ 
		\vspace{6pt}	
		{\frac{{{d^2}z}}{{d{\tau ^2}}} + a{{\left( {\frac{{dt}}{{d\tau }}} \right)}^2} + a{{\left( {\frac{{dz}}{{d\tau }}} \right)}^2} = 0}\\
		{\frac{{{d^2}r}}{{d{\tau ^2}}} = 0} \\ 
\end{array}} \right.
	\right.
\end{equation}
These are parametric equations with respect to $t\left( \tau  \right)$,$z\left( \tau  \right)$ and $r\left( \tau  \right)$, which describe a world line where an object moves freely in Rindler space-time. If we set the initial condition$({t\left( 0 \right) = 0,z\left( 0 \right) = {z_1},r\left( 0 \right) = {r_1}})$, the parametric equations of geodesics can be obtained.
\begin{equation}
	\label{10}
	\left.
\begin{gathered}
	t = \frac{1}{{2a}}\ln \left( {\frac{{\left( {{c_1} + a\tau } \right){c_2}}}{{\left( {{c_2} + a\tau } \right){c_1}}}} \right), \hfill \\
	z = \frac{1}{{2a}}\ln \left( {\frac{{\left( {{c_1} + a\tau } \right)\left( {{c_2} + a\tau } \right)}}{{{c_1}{c_2}}}} \right) + {z_1}, \hfill \\
	r = \frac{{{e^{a{z_1}}}}}{{\sqrt { - {c_1}{c_2}} }}\tau  + {r_1}. \hfill \\ 
\end{gathered} 
	\right.
\end{equation}
${c_1},{\text{ }}{c_2}$ are the integral constants. In order to get the ABCD matrix describing optical paraxial transmission, we need to know the linear transformation relationship between the position and direction of the input field and the output field. 
\begin{figure}[htbp]
	\centering
	\includegraphics[scale=0.32]{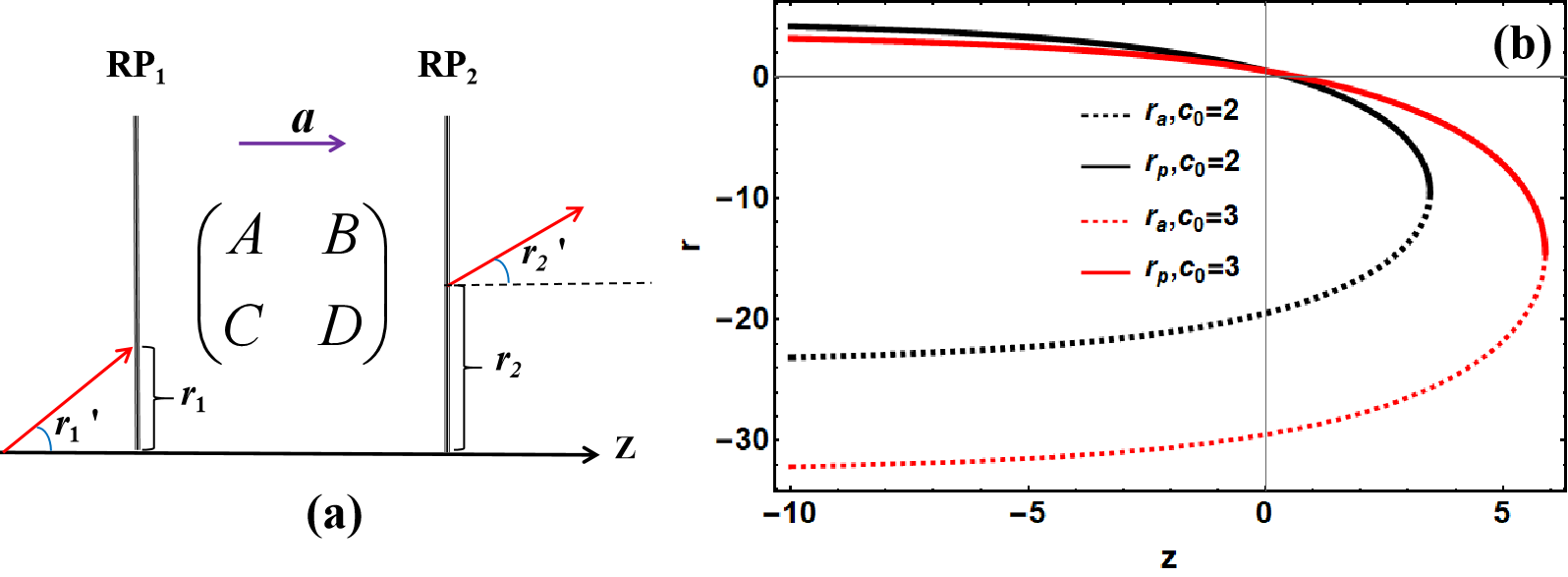}% Here is how to import EPS art
	\caption{\label{1}(a) Accelerated observation ideal experiment diagram, where $RP_1$ and $RP_2$ represent the input plane and output plane, respectively. (b) Ideal trajectories of light rays in different initial directions, dashed lines represent abaxial transmission, while solid lines represent paraxial transmission.}
\end{figure}

By substitution, we can deduce the equation that satisfies the lateral position.
\begin{equation}
	\label{11}
	\left.
	{a^2}{e^{ - 2a{z_1}}}{\left( {r - {r_1}} \right)^2} + {c_0}a{e^{ - a{z_1}}}\left( {r - {r_1}} \right) = 1 - {e^{2a\left( {z - {z_1}} \right)}},
	\right.
\end{equation}
where the integral constant ${c_0} = \sqrt { - {c_1}/{c_2}}  + \sqrt { - {c_2}/{c_1}}$, and its physical significance is related to the reciprocal of the angle in the direction of light transmission in the flat space reference frame:${c_0} = 2dZ/dX$.

Eq. \eqref{11} has two solutions ${r_p}$ and ${r_a}$, corresponding to the paraxial and abaxial solutions in Fig. 1(b), respectively, here we only consider the paraxial solution ${r_p}$. If we define the position and the propagation angle of the field as $r$ and $r' = dr/dz$, shown in Fig. 1(a), by utilizing Taylor expansion and the first-order approximation, we can naturally derive the following transformation relation:
\begin{equation}
	\label{12}
	\left.
\begin{gathered}
	{r_2} = {r_1} + \left( {\frac{{{e^{2a\left( {{z_2} - {z_1}} \right)}} - 1}}{{2a}}} \right){r_1}' + O\left[ {{r_1}{'^3}} \right], \hfill \\
	{r_2}' = {e^{2a\left( {{z_2} - {z_1}} \right)}}{r_1}'+ O\left[ {{r_1}{'^3}} \right]. \hfill \\ 
\end{gathered} 
	\right.
\end{equation}
Following the matrix optics notation and ignoring the higher order terms, the transformation matrix can be deduced.
\begin{equation}
	\label{13}
	\left.
\left( {\begin{array}{*{20}{c}}
		{{r_2}} \\ 
		{{r_2}'} 
\end{array}} \right) = \left( {\begin{array}{*{20}{c}}
		A&B \\ 
		C&D 
\end{array}} \right)\left( {\begin{array}{*{20}{c}}
		{{r_1}} \\ 
		{{r_1}'} 
\end{array}} \right) = \left( {\begin{array}{*{20}{c}}
		1&{\frac{{{e^{2a({z_2} - {z_1})}} - 1}}{{2a}}} \\ 
		0&{{e^{2a\left( {{z_2} - {z_1}} \right)}}} 
\end{array}} \right)\left( {\begin{array}{*{20}{c}}
		{{r_1}} \\ 
		{{r_1}'} 
\end{array}} \right).
	\right.
\end{equation}
A transformation matrix can also be obtained as ${M_{{\text{rd}}}}$ when the acceleration $a$ approaches zero.
\begin{equation}
	\label{14}
	\left.
	{M_{{\text{rd}}}} = \left( {\begin{array}{*{20}{c}}
			A&B \\ 
			C&D 
	\end{array}} \right)\mathop  = \limits^{a \to 0} \left( {\begin{array}{*{20}{c}}
			1&{{z_2} - {z_1}} \\ 
			0&1 
	\end{array}} \right). 
	\right.
\end{equation}
This is consistent with the transformation matrix of a free flat space.

The intrinsic correlation of among equivalent gravity (acceleration system), geometry description and refractive index is interesting for the transmission of beam in space-time background. From the perspective of geometry, if we think about the $r-z$ plane in Rindler space-time at a fixed time, it can be embedded in three-dimensional Euclidean space just like some black hole simulations. Then Eq. \eqref{metri} can be written as
\begin{equation}
\begin{split}
	\label{15}
d{s^2} = {e^{2az}}d{z^2} + d{r^2} = d{z^2} + d{q^2} + d{r^2} \\
= \left( {1 + {{\left( {\frac{{dq}}{{dz}}} \right)}^2}} \right)d{z^2} + d{r^2}.
	\end{split}
\end{equation}
For $a>0$,
\begin{equation}
	\label{16}
	\left.
	q\left( z \right) = c \pm \frac{{\sqrt {{e^{2az}} - 1}  - \arctan \left( {\sqrt {{e^{2az}} - 1} } \right)}}{a}.
	\right.
\end{equation}
The rotational surface$\left( {q\cos \theta ,{\text{ }}q\sin \theta ,{\text{ }}z} \right)$ can be constructed by Eq. \eqref{16}, as shown in Fig. 2, with geometrical properties similar to the $r-z$ frame at a fixed time in Rindler space-time.
\begin{figure}[htbp]
	\centering
	\includegraphics[scale=0.45]{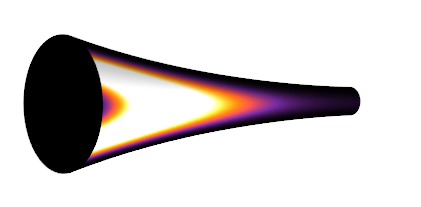}% Here is how to import EPS art
	\caption{\label{2}Surface $\left( {q\cos \theta ,{\text{ }}q\sin \theta ,{\text{ }}z} \right)$ as an analog of Rindler plane.}
\end{figure}

In addition, it is not difficult to find that Eq. \eqref{13} is similar to the optical matrix of light passing through a medium with an axially graded index of refraction $n\left( z \right) = exp\left( { - 2az} \right)$. This is consistent with the equivalent vacuum refractive index of gravity obtained by Ye’s work \cite{RN23}. According to Einstein's principle of equivalence, our equivalent gravitational potential $P\left( z \right) = az$, and the equivalent vacuum refractive index satisfy:\\
\begin{equation}
	\label{17}
	\left.
n\left( z \right) = \exp \left[ { - 2P\left( z \right)/{c^2}} \right],
	\right.
\end{equation}
where we set the speed of light $c = 1$ at the beginning. Of particular note is that although the equivalent refractive index method looks immature, we are still glad to see that our results agree with it. 

It is already verified that the accelerators give us an extra thermal effect \cite{RN18,RN27}, but now we infer that the accelerators could have an “extra refractive index”. The Rindler vacuum can be understood as the variable refractive index medium in Minkowski space-time. This idea goes against general relativity while it has a good use in optics \cite{RN28,RN29,RN30,RN31,RN32}, which is obviously an optical illustration of the vacuum not empty. The difference here is that the “vacuum refractive index” does not affect the speed of light (in this case, light travels at a constant speed), but rather affects the structure of space-time to produce a phenomenon similar to the bending of light, which will be proved in the following.

\begin{table*}[htbp]
	\centering
	\includegraphics[scale=0.5]{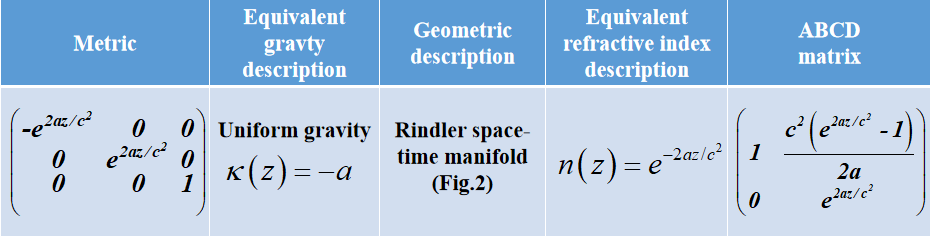}% Here is how to import EPS art
	\caption{\label{2.5}A table of three perspectives with Rindler reference frame. It shows that the equivalent vacuum refractive index has some strong relation with gravity (or geometry). }
\end{table*}

Now, we can deal with the four-dimensional optical transmission in Rindler space-time background simply by using the transformation matrix. 
By diffraction integral theory, the evolution of the light field can be given by the ABCD matrix. The output light field ${E_2}({x_2},{y_2})$, obtained after the input light field ${E_1}({x_1},{y_1})$ passing through the ABCD system, satisfies the Collins formula \cite{RN33}.
\begin{equation}
	\label{18}
	\left.
\begin{gathered}
	{E_2}({x_2},{y_2})\!=\!\left(\!{-\!\frac{{ik}}{{2\pi B}}} \right){e^{ik{L_0}}}\!\times\!\!\!\int\limits_{-\infty }^\infty\!  {\!\!{E_1}({x_1},{y_1}){\text{exp}}\left( {ikL}\right)} d{x_1}\!d{y_1}, \hfill \\
	L = \!\frac{1}{{2B}}\left( {A{{\left( {{x_1} + {y_1}} \right)}^2} - 2\left( {{x_1}{x_{\text{2}}} + {y_1}{y_2}} \right){\text{ + }}D{{\left( {{x_2} + {y_2}} \right)}^2}} \right). \hfill \\ 
\end{gathered}
	\right.
\end{equation}
In this formula, $k = 2\pi /{\lambda _c}$, ${\lambda _c}$ represents the vacuum wavelength in flat space which is constant, but the true wavelength (or frequency) varies during the propagation of light beam. The gravitational redshift is predicted by general relativity:\\
\begin{equation}
	\label{19}
	\left.
	\begin{gathered}
		\lambda \left( {{z_2}} \right)/\lambda \left( {{z_1}} \right) = \sqrt {{g_{11}}\left( {{z_2}} \right)} /\sqrt {{g_{11}}\left( {{z_1}} \right)} {\text{ = }}exp\left[ {a({z_2} - {z_1})} \right], \hfill \\
		f\left( {{z_2}} \right)/f\left( {{z_1}} \right) = \!\sqrt {{ -}{g_{00}}\left( {{z_1}} \right)} /\sqrt {{ -}{g_{00}}\left( {{z_2}} \right)} {\text{ = }}exp\left[ {a({z_1} - {z_2})} \right]. \hfill \\ 
	\end{gathered}
	\right.
\end{equation}
In the Rindler space-time, based on Eq.\eqref{19}, it is obviously that ${\text{c = }}\lambda f$ is constant, which is a common conclusion in GR.
\section{ HGB propagation in accelerated frame}

The research of hollow beam is processing rapidly in recent years \cite{RN34,RN35,RN36,RN37}. Our research is based on the simplest hollow light model and focuses on its spatial distribution because we hope that the transformation of the reference frame can be applied to the beam propagation with some generality.

Under the premise of some reasonable assumptions, we consider the propagation of HGB in an accelerated frame. In general, we set the initial hollow Gaussian beam simply as follows \cite{RN35}:
\begin{equation}
	\label{20}
	\left.
	{E_1}({x_1},{y_1}) = {\left( {\frac{{{x_1}^2 + {y_1}^2}}{{{\sigma ^2}}}} \right)^n}{e^{ - \frac{{{x_1}^2 + {y_1}^2}}{{{\sigma ^2}}}}},
	\right.
\end{equation}
$n$ represents the order of the HGB, $\sigma$ denotes to initial spot size. The hollow Gaussian beam satisfies good hollow characteristic under the near field conditions, but becomes no longer “hollow” in the far field when propagating in the free flat space. Using the treatment described in the first part, Eq.\eqref{20} is substituted into the Collins’ formula, do the integral, and we can get an analytic solution: 
\begin{equation}
	\label{21}
	\left.
\begin{gathered}
	{E_2}\left( {{x_2},{y_2}} \right) = {\left( {\frac{1}{2} - \frac{{iAk{\sigma ^2}}}{{4B}}} \right)^{ - 1 - n}}\frac{{k{\sigma ^2}}}{{2iB}}{{\text{e}}^{\frac{{iDk}}{{2B}}\left( {{x_2}^2 + {y_2}^2} \right) + ik{L_0}}} \\ 
	\times \;\Gamma (1 + n){\text{La}}\left( { - 1 - n,\frac{{ - {k^2}{\sigma ^2}\left( {{x_2}^2 + {y_2}^2} \right)}}{{2B\left( {2B - 4ik{\sigma ^2}} \right)}}} \right). \\ 
\end{gathered}
	\right.
\end{equation}

\begin{figure*}[htp]
	\centering
	\includegraphics[scale=0.34]{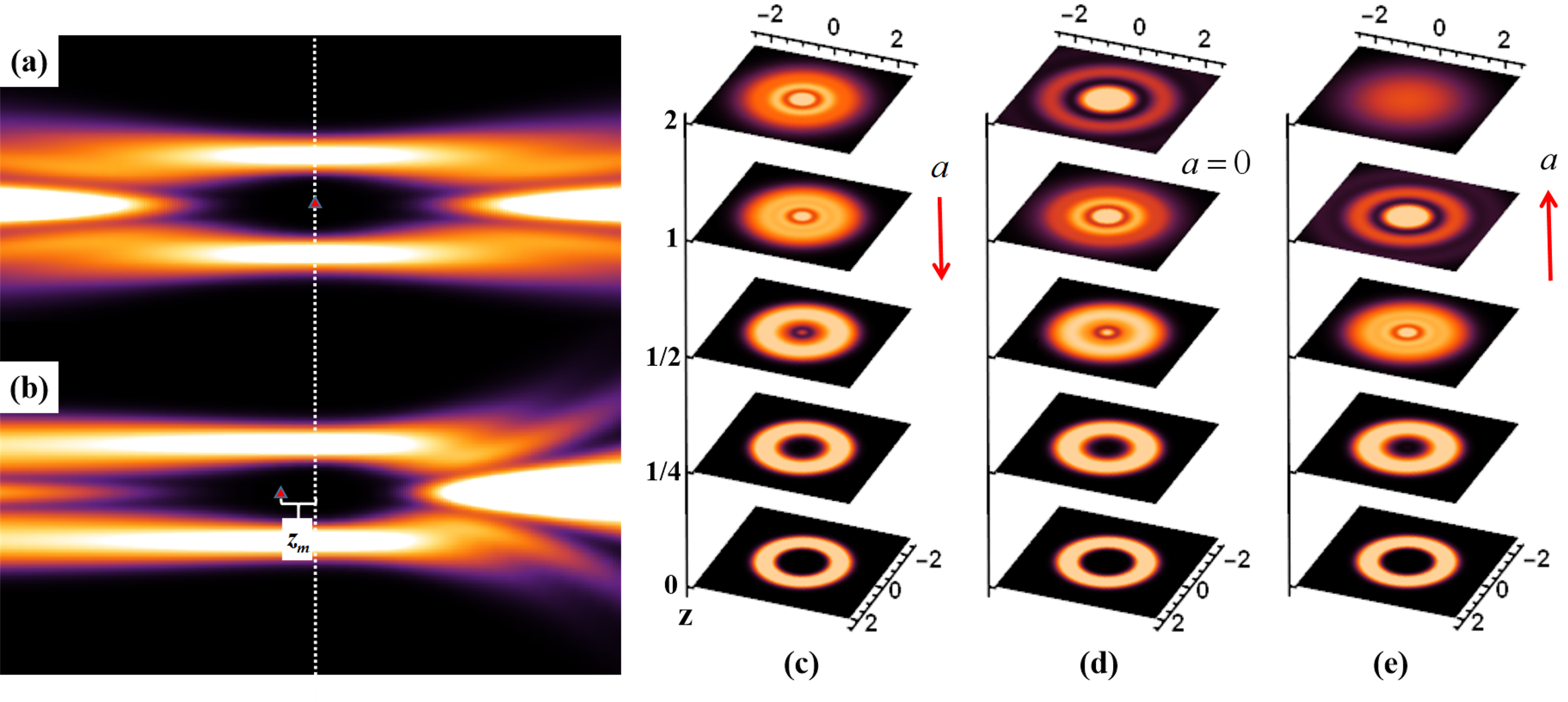}% Here is how to import EPS art
	\caption{\label{3}The intensity distributions of hollow beams in Rindler spacetime (b) (c) (e) and flat Minkowski spacetime(a) (d). The leftmost column (a) and (b) represent the intensity distribution on the $r-z$ plane and the others show the transverse intensity distribution at different propagation plane. The red arrow represents the direction of acceleration.}
\end{figure*}

The ABCD matrix in the formula has been given in Eq. \eqref{13}, $\Gamma$ represents the gamma function and ${\text{La}}(n,z)$ represents the Laguerre polynomials of order $n$ , it degrades to a Gaussian beam when $n=0$. Strictly speaking, the above Eq. \eqref{21} has some defects, such as whether the wave number $k$ is constant in Rindler space-time. The Collins formula, which is derived from the eikonal function, can be widely applied in the medium transmission field. Therefore, when we utilize this formula, it has been assumed that our previous assumption of vacuum refractive index is valid, so we can use the standard diffraction integral. In the figures below, we uniformly set the Rayleigh distance as $1$.

We focus on the spatial distribution of the light intensity of hollow beams during its propagation. As shown in Fig. 3(a)-(b), we demonstrate that in the Rindler frame, the longitudinal dark length and transverse dark width of the beam are different compared to those in the inertial frame. For example, the qualitative dark area in the middle of the hollow beams is an olive shape in flat Minkowski frame, but the intensity symmetry of the coordinate origin along its propagation axis in the Rindler frame is broken. Furthermore, the longitudinal length of the dark area would be stretched and the center of the dark area would shift. In Fig. 3(b) the direction of the acceleration is to the right, so the intensity distribution of hollow beam is compressed in the right part of $z=0$ and stretched in the left part, which lead to the center of the dark area move to the left. Fig. 3(c)-(e) show the transverse intensity distributions in the transmission process for three different accelerated situations. The evolution trends of the central field intensity are similar, increasing from zero and then decreasing; but the forward acceleration can accelerate such evolution process, otherwise it will slow down.

Fig. 4(a) shows the distribution of light intensity along the axis of propagation for different accelerated value. For hollow beams, there are two polar points in the direction of propagation as 
\begin{equation}
	\label{22}
	\left.
\frac{{d{{\left| {{E_2}} \right|}^2}_{\left( {0,0} \right)}}}{{dz}} = 0.
	\right.
\end{equation}
By solving the Eq. \eqref{22},we have the value of two polar points,
\begin{equation}
	\label{23}
	\left.
	{z_f} = - \frac{1}{{2a}}\ln \left( {\frac{1}{{1 + a{z_0}}}} \right),{\text{  }}{z_b} =  - \frac{1}{{2a}}\ln \left( {\frac{1}{{1 - a{z_0}}}} \right),
	\right.
\end{equation}
where ${z_0} = \sqrt n k{\sigma ^2}$ is the length of the dark spot in an acceleration free inertial frame.
\begin{figure*}[htb]
	\centering
	\includegraphics[scale=0.3]{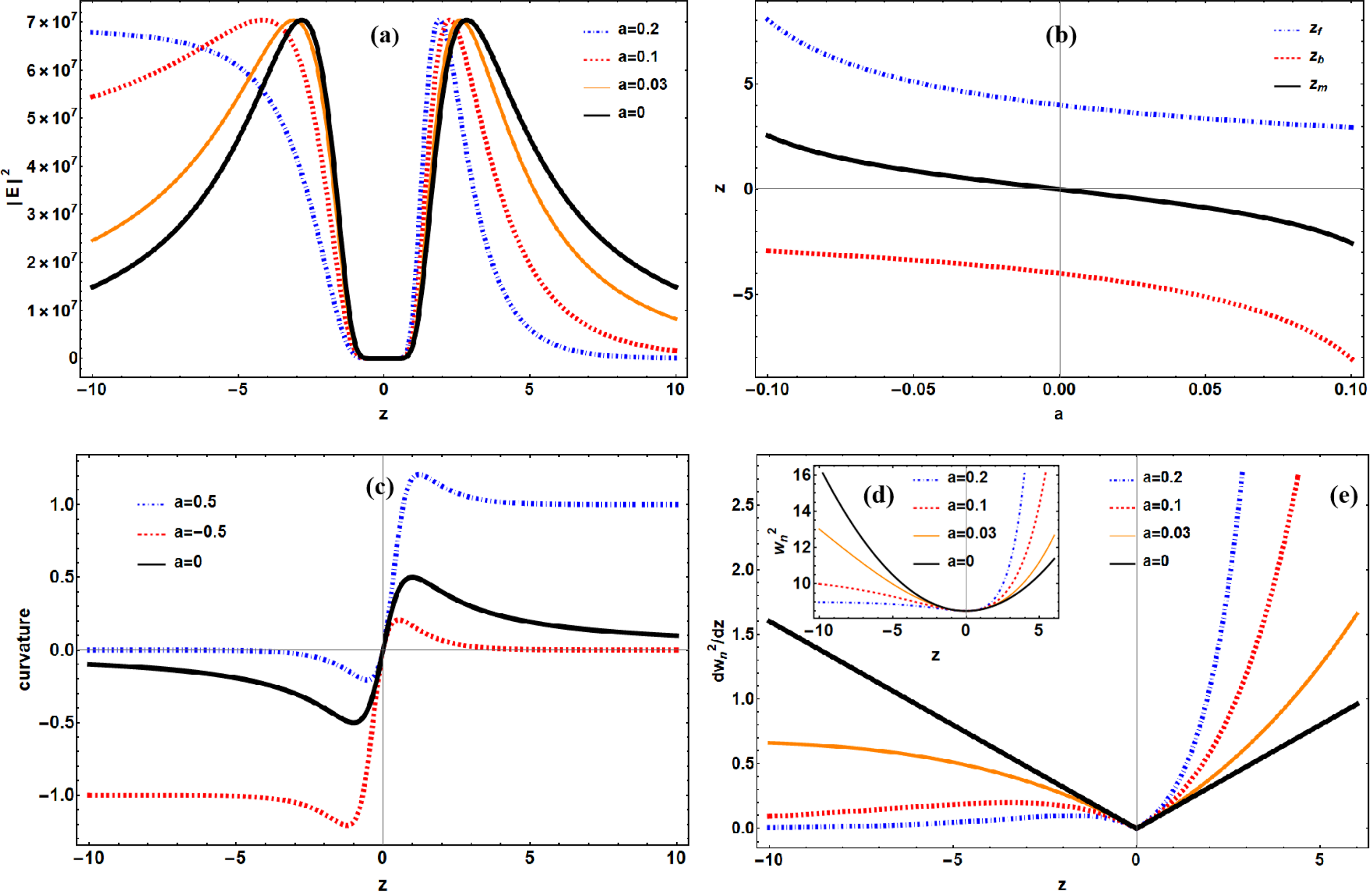}% Here is how to import EPS art
	\caption{\label{4.5}The evolution of parameters in the Rindler reference. (a) The distribution of beam intensity along the axis at different accelerations; (b) Evolution of the dark spot boundary and center with the reference frame acceleration. (c) Propagation of wavefront curvature of a beam; The spot area (d) and the change rate of spot area (e) with propagation.}
\end{figure*}
Obviously, ${z_f} =  - {z_b} = {z_0}/2$ as $a$ approaches $0$, that is, the symmetry propagation is satisfied in inertial frame, but the equivalent gravitational field will shift and stretch the hollow region. In the non-inertial reference frame, the center of dark spot shifts towards the opposite direction of the acceleration as Fig. 4(b). We can use the following two formulas to approximate the shift ${z_m}$ and the stretch $\Delta {z_s}$.
\begin{equation}
	\label{24}
	\left.
	{z_m} = \frac{{{z_f} + {z_b}}}{2} \approx  - \frac{1}{4}n{k^2}{\sigma ^4}a,
	\right.
\end{equation}
\begin{equation}
	\label{25}
	\left.
\Delta {z_s} = \left( {{z_f} - {z_b}} \right) - {z_0} \approx \frac{1}{3}{n^{3/2}}{k^3}{\sigma ^6}{a^2}.
	\right.
\end{equation}

Eq. \eqref{24} and Eq. \eqref{25} mean that the gravitational field can change the hollow region of a hollow beam, which gives a new potential to gravity measurements or the accelerator optics. 

The equivalent gravitational field also accelerates the bending of the beam's wavefront towards the opposite direction of the coordinate acceleration. Because different geodesics lead to distinct real values and imaginary values of phase delay, the field intensity distribution and wavefront curvature of the beam are affected accordingly. Fig. 4(c) shows the change of wavefront curvature. The most significant difference compared with the inertial system is that the wavefront in the positive direction of acceleration cannot be regarded as a plane wave even after propagating a quite far distance, but a fixed spherical wave. As the wavefront curvature does not converge to 0 but a constant at the far field as Eq. \eqref{26}.
\begin{equation}
	\label{26}
	\left.
\frac{1}{{{\rho _{{\text{z}} \to \infty }}}} = 2a
	\right.
\end{equation}
We think with the increasing of the velocity due to the acceleration $a$, the maximum velocity approaches the propagation speed of light in vacuum, then the product of $az$ becomes a constant. That is the reason why the wavefront curvature is a constant at the far field.
The higher-order HGB will be split into multiple rays during propagation, we can calculate the transverse beam width in the average sense with the propagation as follows:
\begin{equation}
	\label{27}
	\left.
	\!{w_n}^2\!=\!\frac{{\iint\limits_{x,y} {{{\left| {{E_2}} \right|}^2}\left( \!{{x^2}\!+\!{y^2}}\right)\!dxdy}}}{{\iint\limits_{x,y} {{{\left| {\!{E_2}} \right|}^2}dxdy}}}\!=\!\frac{{{{\left( {{e^{2az}}\!-1} \right)}^2}}}{{2{k^2}{\sigma ^2}{a^2}}}\!+\!\frac{{\left( {2n \!+\! 1} \right)}}{2}{\sigma ^2}.
	\right.
\end{equation}

In the case of non-inertia, the transverse spot of the beam shows obvious symmetry. Its divergence in the direction of acceleration is faster than that in the inertial frame, which is anti-focusing; while in the opposite direction, this process is slower than the inertial frame, appearing the focusing effect as Fig. 4(d). It is worth mentioning that the transverse beam width at $z \to  - \infty$ plane is finite.
\begin{equation}
	\label{28}
	\left.
w_n^2(z \to  - \infty ) = \frac{\pi }{{2{k^2}{\sigma ^2}{a^2}}} + \frac{{2n + 1}}{2}\pi {\sigma ^2}.
	\right.
\end{equation}
This is not the same as the transmission in flat space, which diverges to infinity as the propagation distance increases. From Fig. 4(e), when propagating backwards compared to the acceleration, the growth rate of beam size in the Rindler reference frame has the maximum value and then converges to 0. Therefore, we can say that the attraction effect of equivalent gravity on the light itself constrains the light's divergence. The extreme point position $z_i$ is elegant, which is inversely proportional to the acceleration.
\begin{equation}
	\label{29}
	\left.
a{z_i} =  - \frac{{\ln 2}}{2}.
	\right.
\end{equation}

We further consider that the wave properties of light imply uncertainty for photons.The size of the transverse beam spot can be utilized to describe the coordinate uncertainty of the photon on the level of quantum mechanics. In the inertial frame, the position uncertainty of the photon increases linearly with the evolution of time, while Eq. \eqref{27} tells us that the accelerated frame can change the linear evolution of the coordinate uncertainty of the photon. When we consider forward transport of Gaussian beam with minimum uncertainty($n=0$), we can estimate the uncertainty of transverse position $\Delta x$ and momentum $\Delta p$.
\begin{equation}
	\label{30}
	\left.
\begin{gathered}
	{\left( {\Delta x} \right)^2} = {w_0}^2\left( {z = 0} \right) = \frac{1}{2}{\sigma ^2}, \hfill \\
	{\left( {\Delta p} \right)^2} = {\left( {p\Delta \theta } \right)^2} \approx {p^2}\frac{{{w_0}^2 - {w_0}^2\left( {z = 0} \right)}}{{{z^2}}}. \hfill \\ 
\end{gathered} 
	\right.
\end{equation}
Introducing the de Broglie hypothesis $p = \hbar k$, the uncertainty relationship must be different from the flat space case as Eq. \eqref{31}, and this has been proved by theoretical physicists in the quantum field theory (GUP) \cite{RN19}.
\begin{equation}
	\label{31}
	\left.
{\left( {\Delta x} \right)^2}{\left( {\Delta p} \right)^2} \sim {\hbar ^2}\left( {1 + 2az + O\left[ {{a^2}{z^2}} \right]} \right).
	\right.
\end{equation}
It is definitely encouraging for our calculations.
\section{Conclusion}
General relativity states that the gravitational field and the accelerating field are locally equivalent, and we study the propagation of light beams through such a vacuum-- a vacuum that is uniformly accelerated or in a uniformly constant gravitational field. The geometrical properties of light imply that the beam propagation is closely related to the space-time background. Starting from these assumptions, we investigate the relationship between Rindler space-time and the medium with variable refractive index.

Based on the wave equation and geodesic equations, the ABCD transfer matrix of Rindler vacuum system is deduced. The high similarity of its form to the variable refractive index medium suggests that these two are closely related in a sense. By utilizing a Collins formula, we analyze the transmission properties of HGB in Rindler space-time frame, including the shift and stretch of the axial dark spots, convergence of transverse beam size as well as a constant wavefront curvature. We also explore the necessity to modify the Heisenberg uncertainty relation in Rindler frame. It is hoped that we provide a simple calculation model for optical control and action of accelerated particles.

%\begin{acknowledgments}
%
%\end{acknowledgments}
%
%
%
%\section{Appendixes}


%\bibliography{reference}% Produces the bibliography via BibTeX.


\end{document}